\documentclass{jfm}
\usepackage{graphicx}
\usepackage{epstopdf, epsfig}
\usepackage{xcolor}

\shorttitle{Airfoil Synchronous Surging and Pitching}
\shortauthor{Strangfeld et al.}

\title{Airfoil Synchronous Surging and Pitching}

\author{C. Strangfeld,\aff{1}\aff{3}
  H.F. M{\"u}ller-Vahl,\aff{2}
  C.N. Nayeri,\aff{3}
  C.O. Paschereit\aff{3}
 \and D. Greenblatt\aff{2}\corresp{\email{davidg@technion.ac.il}}}

\affiliation{$^1$Bundesanstalt f{\"u}r Materialforschung und -pr{\"u}fung, Unter den Eichen 87, 12005 Berlin, Germany\\[\affilskip]
$^2$Faculty of Mechanical Engineering, Technion - Israel Institute of Technology, Haifa 3200003, Israel\\[\affilskip]
$^3$Hermann-F{\"o}ttinger Institut, Institute of Fluid Dynamics and Technical Acoustics, Technische Universit{\"a}t Berlin, M{\"u}ller-Breslau-Str. 8, 10623 Berlin, Germany}

\begin{document}

\maketitle

\begin{abstract}
Combined pitching-and-surging of an airfoil at the identical frequency (i.e., synchronously), at four different phase-differences, was investigated theoretically and experimentally.  The most general unsteady theoretical formulation, developed by \cite{vanderwall1992forschungsbericht}, was adopted to calculate the lift coefficient, which has both circulatory and non-circulatory components. Unsteady airfoil theory was further extended to explicitly compute the unsteady bound vortex sheet, which could be directly compared to experiments and facilitated the computation of both Joukowsky and impulsive-pressure lift contributions. Experiments were performed, employing a NACA 0018 airfoil, in an unsteady wind tunnel at an average Reynolds number of $3.0\times 10^5$, with a free-stream oscillation amplitude of 51~\%, an angle-of-attack range of $2^\circ \pm 2^\circ$, and a reduced frequency of 0.097. In general, excellent correspondence was observed between theory and experiment, representing the first direct experimental validation of the general theory. It was shown, both theoretically and experimentally, that the lift coefficient was not accurately represented by independent superposition of surging and pitching effects, due to variations in the instantaneous effective reduced frequency not accounted for during pure pitching. Deviations from theory, observed at phase-lags of $90^\circ$ and $180^\circ$ were attributed to bursting of separation bubbles during the early stages of the acceleration phase. The largest deviations occurred when the impulsive-pressure lift contribution was small relative to the Joukowsky contribution, because the latter was most affected by bubble-bursting. Bubble-bursting resulted in large form-drag oscillations that occurred at identical phase angles within the oscillation cycle, irrespective of the phase difference between surging and pitching, as well as in the absence of pitching. The bubble-bursting dynamic stall mechanism, observed here at low pre-stall angles of attack, may have important implications for rotor performance and noise emissions.
\end{abstract}


\section{Introduction}\label{sec:motivation}
Low-speed unsteady aerodynamics is concerned with determining the unsteady loads on fixed wing aircraft \citep[e.g.,][]{bisplinghoff2013aeroelasticity,jones2022physics}, rotorcraft \citep[e.g.,][]{sharma2019numerical, gardner2023review}, wind turbines \citep[e.g.,][]{ferreira2009visualization, simms2001nrel, buchner2018dynamic} and flapping-wing flyers \citep[e.g.,][]{taylor2003flying}. The majority of idealized studies involve two-dimensional airfoils undergoing dynamic rotation or translation, most commonly by periodic pitching or plunging \citep[see references in  the reviews by][] {corke2015dynamic,turhan2022coherence, bergami2013indicial}. During the last decade, streamwise (or longitudinal) oscillation of the flow or test article---often called surging---has received increased attention \citep[e.g.,][]{dunne2015dynamic, granlund2014airfoil, choi2015surging, medina2018highamplitude, muellervahl2020dynamic}, due to its relevance to rotorcraft and wind turbines. Surging studies present the technical challenge of attaining large relative amplitudes $(\sigma \equiv \Delta u /u_s \gtrsim 20~\%)$, at practically relevant reduced frequencies $k \equiv \omega c/ 2 u_s$, and several different approaches have been proposed \citep[e.g.,][]{granlund2014airfoil,greenblatt2001unsteady,gloutak2024aerodynamic}. Here  $u_s$ is the cycle-averaged free-stream velocity, $\omega$ is the circular frequency and $c$ is the chord-length. In contrast to pitching and plunging, surging can introduce significant time-dependent Reynolds number effects \citep[e.g.,][]{carmichael1981low,toppings2023transient}, typically observed when $Re(t) \equiv u(t) c/\nu \lesssim 10^{5}$, where $u(t)$ is the time dependent free-stream velocity and $\nu = \mu / \rho$ is the kinematic viscosity.

Recently, investigators have recognized the importance of studying surging at nominally pre-stall angles of attack \citep{strangfeld2016airfoil,yang2017experimental, zhu2020unsteady, duncan2024powered, gloutak2024aerodynamic, wang2024airfoil}. Two of these experiments considered relatively high $\{{Re_s},\sigma \}$ NACA 0018 experiments \citep[][the first one being or own]{strangfeld2016airfoil, zhu2020unsteady} in the range $0{}^\circ \le \alpha \le 4{}^\circ$. With an aerodynamically smooth low-pressure airfoil surface, the phase-averaged lift coefficient ${{C}_{l}}(\phi)$ in our investigation showed an \emph{anti-correlation} with the theories of \cite{greenberg1947airfoil} and \cite{isaacs1945airfoil}. However, with the boundary layer passively perturbed by a two-dimensional discontinuity (slot) at $x/c=5~\%$, the lift coefficient corresponded with theory, but high-frequency oscillations were observed during the early stages of the acceleration phase.  In a follow-up study \citep{greenblatt2023laminar}, we identified the high-frequency oscillations as laminar separation bubble (LSB) bursting on both surfaces of the airfoil, during the early stages of the \emph{favorable temporal pressure gradient}. It was hypothesized that the stabilizing effect of the slot perturbation on the LSB acts to delay or prevent its bursting \citep{yarusevych2017steady, marxen2011effect} and, therefore, this was the most likely reason for the improved correspondence with theory. Bubble-bursting also manifested as large differences between corresponding unsteady and quasi-steady form-drag coefficients ${{C}_{d}}(\phi)$ and ${{C}_{d,{qs}}}(\phi)$.  Phase-averaged lift coefficient ${{C}_{l}}(\phi)$ data of \cite{zhu2020unsteady} showed significant deviations from the theory of \cite{isaacs1945airfoil}. In particular, under the conditions $\sigma = 0.23$ and $k = 0.050$, the measured lift coefficient exceeded the theoretical value by an order of magnitude with multiple local lift peaks. Using a technique known as background-oriented Schlieren (BOS) visualization, they demonstrated that the classical Kutta condition is violated under surging. This was attributed to either a pressure difference across trailing-edge due to viscosity \citep[based on the analysis of][]{taha2019viscous}, or trailing-edge separation. The trailing-edge separation hypothesis advanced by \cite{zhu2020unsteady} is consistent with our own observations of bubble-bursting.

The combined effect of simultaneous surging and pitching on airfoil loads is a scenario that has greater direct practical relevance, particularly as an idealized representation of rotorcraft and wind turbine aerodynamics. The majority of these studies   \citep[e.g.,][]{dunne2015dynamic, granlund2014airfoil, choi2015surging}, including our own \citep{medina2018highamplitude, muller2017matched,muellervahl2020dynamic},  focus on the phenomenon of dynamic stall observed at post-stall angles-of-attack $\alpha >{{\alpha}_{s}}$, due to its technological importance. One exception is the experimental NACA 0015 airfoil study of \cite{ma2021unsteady}, where small pitch amplitude oscillations of $\Delta \alpha = 1^{\circ}$ and $2^{\circ}$ around zero in the presence of surging with $\sigma=0.2$, was considered. The pitching and surging frequencies were different, with no phase relationship between them, and the data showed similar trends to the theory of \cite{greenberg1947airfoil}. In contrast to the study of \cite{ma2021unsteady}, idealized rotorcraft and wind turbine blades experience simultaneous, or synchronous, surging-and-pitching, i.e., the frequency of both is the same although they are generally not in phase. To date, no direct experimental validation of the general synchronized surging-and-pitching theory has been conducted. Furthermore, explicit computation of the unsteady bound vortex sheet, that distinguishes between so-called ``Joukowsky'' lift and ``impulsive-pressure'' lift has not been performed.

The objective of the present research, therefore, is to study, both theoretically and experimentally, the effects of  synchronous surging-and-pitching on an airfoil at low pre-stall angles-of-attack $(0{}^\circ \le \alpha \le 4{}^\circ )$. The theoretical approach follows the methods developed by \cite{vanderwall1994on} for synchronous surging-and-pitching, and then extends them to calculate the bound, unsteady vortex sheet strength. In the experiments, we use the tripped low-surface pressure cases that corresponded with theory under pure surging (i.e., surging at constant $\alpha$), and add synchronous pitching at four different phase-differences, namely $\tau = 0^{\circ}$, $90^{\circ}$, $180^{\circ}$ and $270^{\circ}$.  This facilitates a comparison of the unsteady lift coefficients. Much like our pure surging study, comparisons of the measured and theoretical vortex sheet strengths are used to determine the limits of the theory and, in particular, the impact of separation bubble bursting. Furthermore, we directly compare unsteady surface pressure coefficients with their quasi-steady counterparts in order to understand the bubble-bursting mechanism and its effect on the form-drag coefficient.

\section{Surging \& Pitching Airfoil Theory}\label{sec:theory}
\subsection{Unsteady Lift Overshoot in Potential Flow}
\begin{figure}
\centering
\includegraphics[width=380px]{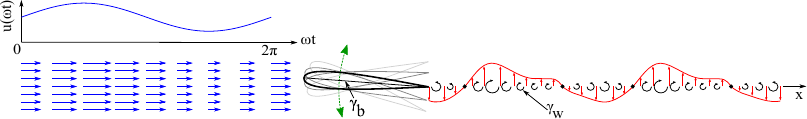}
\caption{Schematic illustrating a pitching airfoil in a surging  free-stream, which generates an unsteady wake vortex sheet.}
\label{fig:wake} 
\end{figure}

The problem considered in this paper is illustrated schematically in Figure \ref{fig:wake}, that shows an airfoil (in our case, a NACA 0018 airfoil) that is  pitched harmonically about its quarter-chord position (green arrow), while the free-stream surges harmonically at the identical frequency. The length of the blue arrows schematically illustrates the time varying velocity amplitude and the phase-difference between the free-stream and angle-of-attack is assigned the symbol $\tau$. The lift force acting on the airfoil at any phase-angle $\phi$, which is proportional to the circulation of the bound vortex sheet, is affected by a combination of the unsteady inflow and airfoil motion. According to Helmholtz's circulation theorem, the overall circulation in the global system must remain constant. Thus, a circulation change of arbitrary strength at one time step requires the shedding of a vortex into the wake with opposite strength at this time step (illustrated by the black wake vortices and red line). The shed wake vorticity induces normal velocities on the airfoil which modify the circulation change and hence the generated lift. The higher the velocity amplitudes $\sigma$ or the reduced frequencies $k$, the larger the influence of the wake vorticity.\\
Theories for unsteady airfoil lift and pitching moment, first introduced by  \cite{theodorsen1935general}, all invoke the assumption that the airfoil is modelled as a two-dimensional (infinite-span) flat plate in an incompressible potential flow. The boundary layer is modelled as a vortex sheet, approximating a thin airfoil at high Reynolds number under fully-attached flow, and hence no explicit friction forces, diffusion or separation exist. Throughout the derivation of the closed form solution presented here, only small angles-of-attack are considered. In the case of an unsteady, sinusoidal free-stream, the maximum amplitude of the velocity oscillation is limited to $\sigma\leq1$ to prohibit reverse flow.

On the one hand, the theory of \citet{theodorsen1935general} is used to predict the unsteady lift and pitch moment due to various harmonic airfoil motions like pitching, plunging, or flap deflections with a constant free-stream. On the other hand, the theory of \citet{isaacs1945airfoil} computes the unsteady lift and pitching moment of an airfoil at a constant angle-of-attack subjected to a harmonically varying free-stream. The theory of \citet{greenberg1947airfoil} considers the same problem but due to the so-called high frequency assumption, this theory is limited to relatively small velocity ratios $\sigma$ below 0.4 \citep{vanderwall1994on, strangfeld2014airfoil}. However, Isaacs extended his theory to incorporate both degrees of freedom, i.e., surging and mid-chord pitching, simultaneously \citep{isaacs1946airfoil}. Based on this more general analysis, \citep{vanderwall1992forschungsbericht} further extended the theory to include an arbitrary pitch axis, the integration of arbitrary harmonic pitch profiles, and arbitrary vertical airfoil motion. This approach is the most general formulation, and hence it is discussed below in more detail.

\subsection{Theoretical Approach of van der Wall}
Consider an airfoil submerged in an incompressible free-stream velocity, that is surging according to:
\begin{eqnarray}
u(\phi)=u_s(1+\sigma\sin(\phi))
\label{equa:surging}
\end{eqnarray}


\noindent and whose angle-of-attack is oscillating according to:
\begin{eqnarray}
\alpha(\phi)=\alpha_0 \left[\bar{\alpha}_0+\sum_{n=1}^\infty[\bar{\alpha}_{nS}\sin(n\phi)+\bar{\alpha}_{nC}\cos(n\phi)]\right]
\label{equa:pitching}
\end{eqnarray}

\noindent where $\alpha_0$ is the cycle-averaged angle-of-attack, the $\bar{\alpha}$ terms are dimensionless amplitude coefficients and $\phi = \omega t$. The frequency of the free-stream and pitch oscillations $\omega$ are identical, hence only one global $k$, based on the average free-stream velocity ${u}_s$, is used. According to \cite{vanderwall1992forschungsbericht}, the unsteady-to-quasi-steady lift coefficient ratio as a function of $\phi$ is:
\begin{eqnarray}
\frac{C_l(\phi)}{C_{l,qs}} &=& \frac{1}{(1+\sigma\sin(\phi))^2} 0.5k [ ( \sigma\bar{\alpha}_0 +\bar{\alpha}_{1S} + k (a\bar{\alpha}_{1C})-0.5\sigma\bar{\alpha}_{2C} )\cos(\phi) \nonumber\\
&&+(-\bar{\alpha}_{1C}+k(a\bar{\alpha}_{1S}-0.5\sigma\bar{\alpha}_{2S})\sin(\phi) \nonumber\\
&&+\sum_{n=2}^\infty n (\bar{\alpha}_{nS}+nka\bar{\alpha}_{nC}+0.5\sigma(\bar{\alpha}_{(n-1)C}-\bar{\alpha}_{(n+1)C}))\cos(n\phi) \nonumber\\
&&+\sum_{n=2}^\infty n (-\bar{\alpha}_{nC}+nka\bar{\alpha}_{nS}+0.5\sigma(\bar{\alpha}_{(n-1)S}-\bar{\alpha}_{(n+1)S}))\sin(n\phi) ]\nonumber\\
&&+\frac{1}{(1+\sigma\sin(\phi))^2}[( (1+0.5\sigma^2)\bar{\alpha}_0+\sigma(\bar{\alpha}_{1S}-0.5k((0.5-a)\bar{\alpha}_{1C})-0.25\sigma\bar{\alpha}_{2C}))\nonumber\\
&& \cdot(1+\sigma\sin(\phi)) + \sum_{m=1}^\infty (\Re(l_m)\cos(m\phi)+\Im(l_m)\sin(m\phi))]\label{equa:clvanderwall}
\end{eqnarray}

\noindent where the normalised distance of the quarter-chord pitch axis to the midchord corresponds to $a=-0.5$. The first four lines in equation \ref{equa:clvanderwall} express the non-circulatory part of the unsteady lift effects and the last two lines  express the circulatory part that include a summation from the first wave number $m=1$ to infinity. Equation \ref{equa:clvanderwall} furthermore requires the real $\Re$ and imaginary  $\Im$ part of $l_m$, where: 
\begin{eqnarray}
l_m &=& -2m(i)^{-m}\sum^{\infty}_{n=1}[F_n(J_{n+m}(n\sigma)-J_{n-m}(n\sigma))\nonumber\\
&&+iG_n(J_{n+m}(n\sigma)+J_{n-m}(n\sigma))] \label{equa:l_n2}\\
F_n+iG_n&=&[C(nk)]n^{-2}(H_n(n\sigma)+iH_n^\prime(n\sigma))
\end{eqnarray}

The most general formulation of these coefficients is given by  \cite{vanderwall1992forschungsbericht}, which includes Bessel functions of the first kind $J$, the \cite{theodorsen1935general} function $C(nk)=F(nk)+iG(nk)$, and:
\begin{eqnarray}
H_n(n\sigma)&=&\frac{J_{n+1}(n\sigma)-J_{n-1}(n\sigma)}{2}\left[\sigma\bar{\alpha}_0-\bar{\alpha}_{1s}-k(0.5-a)\bar{\alpha}_{1c}\right]-\frac{2J_n(n\sigma)}{n\sigma}\bar{\alpha}_{1s}\label{equa:H_m}\\
H_n^\prime(n\sigma)&=&\frac{J_{n+1}(n\sigma)-J_{n-1}(n\sigma)}{n}\bar{\alpha}_{1c}+\frac{J_n(n\sigma)}{\sigma}\left[\bar{\alpha}_{1c}(1-\sigma^2)-k(0.5-a)\bar{\alpha}_{1s}\right]\label{equa:H_m'}
\end{eqnarray}

The formulation of the coefficients $H_n$ and $H_n^\prime$ in the equations \ref{equa:H_m} and \ref{equa:H_m'} implicitly assumes angle-of-attack oscillations of the form expressed in equation \ref{equa:pitching}. A more general formulation for arbitrary oscillatory motions can be found in \citet{vanderwall1992forschungsbericht}. 

If a constant angle-of-attack is assumed, equation \ref{equa:clvanderwall} reduces to the formulation of \citet{isaacs1945airfoil}. Similarly, if the amplitude of the free-stream velocity oscillation is zero, equation \ref{equa:clvanderwall} is equivalent to the formulation of \citep{theodorsen1935general}. This fortunately obviates the need for a separate presentation of these theories.  Furthermore, equation \ref{equa:clvanderwall} depicts the lift coefficient ratio with $C_l(\phi)/C_{l,qs} = L(\phi)/{L_{qs}}(1+\sigma\sin(\phi))^{-2}$, where the quasi-steady lift $L_{qs}=\pi\rho c u_s^2\alpha $ is determined from steady thin airfoil theory \citep{anderson2011fundamentals}. The ratio of lift coefficients, as opposed to the ratio of lift forces,  is employed here because it clearly shows the net unsteady effects that can otherwise be overwhelmed by the oscillations in dynamic pressure.

Finally, in order explicitly include the phase shift $\tau$ in the formulation, we let $\alpha_0=\alpha_s$, $\bar{\alpha}_0=1$, $\alpha_{1S}=\alpha_{a}\cos(\tau)\alpha_s^{-1}$, and $\alpha_{1C}=\alpha_{a}\sin(\tau)\alpha_s^{-1}$ in equation \ref{equa:pitching}, which can be written as:
\begin{eqnarray}
\alpha(\phi)= \alpha_s+\alpha_{a} [\sin(\phi)\cos(\tau)+\cos(\phi)\sin(\tau)] = \alpha_s+\alpha_{a}\sin(\phi+\tau)
\end{eqnarray}

\subsection{Derivation of the Shed Wake Vorticity}
In the previous section, the phase dependent lift coefficient ratio due to surging-and-pitching was presented. While the this ratio is useful for identifying integral unsteady effects, it does not reveal the local circulation distribution, or vortex sheet strength, along the chord. This distribution is important for two reasons. First, it can be measured in experiments and hence used directly for validation of the theory. Second, deviations from the theory are useful for identifying effects of boundary layer separation. Thus, we seek a new expression for the unsteady bound vortex sheet $\gamma_b(x,t)$ with $x=\frac{c}{2}\cos\Theta$. The first step in this process is the calculation of the time-varying shed wake vorticity strength, which results in induced chord-normal velocities that play a crucial role in determining the unsteady lift variation. The normal velocity distribution along the chord, based on \cite{vanderwall1992forschungsbericht} for small $\alpha$ can be expressed as:
\begin{eqnarray}
v_{n,b}(x,t) &=& \alpha(t)u(t)+(x-0.5ac){\dot{\alpha}}(t)+{\dot{h}}(t)+v_{n,w}(x,t) \label{equa:basicvn}
\end{eqnarray}
The first term includes both the unsteady angle-of-attack as well as the unsteady free-stream velocity, the second term enables an arbitrary positioning of the pitch axis relative to the mid-chord, the third term defines a time-varying vertical airfoil motion and the last term depicts the contribution of the shed vorticity in the wake.

The shed vorticity is generated continuously at the trailing edge and convects downstream, and therefore the shed wake vorticity strength is the time derivative of the unsteady bound circulation, designated $\dot{\Gamma}(\tau^*)$. These shed wake vortices induce velocity components normal to the chord as described by Biot-Savart's law \citep{schade2007stromungslehre} and thus at each instant in time $\tau^*$, the entire unsteady wake needs to be taken into account to determine valid unsteady effects. The wake-induced normal velocities can thus be expressed by:
\begin{eqnarray}
v_{n,w}(\Theta,t) &=& -\frac{1}{2\pi}\int_{-\infty}^t \frac{\dot{\Gamma}(\tau^*)}{1+\frac{W(t)-W(\tau^*)}{0.5c}-\cos(\Theta)}d\tau^* \label{equa:vnw}
\end{eqnarray}
\noindent where $W(t)$ is the distance travelled by the airfoil with the coordinate transformation $x =0.5c\cos(\Theta)$) \citep{vanderwall1992forschungsbericht}.  Normalisation by the half-chord $c/2$ and transformation into a Fourier series yields $v_{n,w}(\Theta,t) = b_0(t)/2+\sum^\infty_{n=1}b_n(t)\cos(n\Theta)$. Furthermore, the self-induced normal velocities $v_{n,b}$ caused by the bound vorticity sheet $\gamma_b$ are given by:
\begin{eqnarray}
v_{n,b}(\Theta,t) &=& \frac{1}{2\pi}\int_0^\pi \frac{\gamma_b(\theta,t)\sin\theta}{\cos(\Theta)-\cos(\theta)} d\theta \label{equa:vnb}
\end{eqnarray}
and its transformation into a Fourier series gives $v_{n,b}(\Theta,t) = d_0(t)/2 +\sum^\infty_{n=1}d_n(t)\cos(n\Theta)$. Now, all quantities in equation \ref{equa:basicvn} are known except for the vorticity sheet $\gamma_b$ which gives the circulation by integration along the chord $\Gamma(t)=\int^{c/2}_{-c/2}\gamma_b(x,t)dx$. The condition of flow tangency is required to solve this problem, and thus the condition:
\begin{eqnarray}
v_n(x,t) &\equiv& 0 = \alpha(t)u(t)+(x-0.5ac)\dot{\alpha}(t)+v_{n,w}(x,t)-v_{n,b}(x,t) \label{equa:tangency}
\end{eqnarray}
\noindent is enforced (see equation \ref{equa:basicvn}) where vertical plunging motion is not considered. A comparison of the Fourier series coefficients yields the following identities \citep{isaacs1945airfoil}:
\begin{eqnarray}
c_0(t)&=& 2\alpha(t) u(t)+c(0.5-a)\dot{\alpha}(t)+ b_1(t) +b_0(t)\label{equa:c0}\\
c_1(t)&=& -2\alpha(t) u(t)+ac\dot{\alpha}(t)+b_2(t)-b_0(t)\\
c_2(t) &=&-0.5c\dot{\alpha}(t)+b_3(t)-b_1(t)\\
c_n(t) &=&b_{n+1}(t)-b_{n-1}(t)~~~~~~~~~~~~~~~~~~~~n\geq3\label{equa:cn}
\end{eqnarray}
Furthermore, the Kutta condition requires finite velocities at the wing's trailing edge $c_0(t)= -\sum^\infty_{n=1}c_n(t)$ \citep{amiet1990gust} and $c_0(t)$ satisfies this condition, because $b_n$ converges to zero for $n\rightarrow\infty$. Finally, the circulation is expressed in cylindrical coordinates according to \citet{isaacs1945airfoil} as
\begin{eqnarray}
\Gamma(t)&=&\frac{c}{2}\int^{\pi}_{0}\gamma_b(\Theta,t)\sin(\Theta) d\Theta\nonumber\\
&=&\frac{c}{2}\int^{\pi}_{0}\left[c_0(t)+\sum^\infty_{n=1}c_n(t)\cos(n\Theta)\right]d\Theta=\frac{\pi c}{2}c_0(t)\label{equa:Gamma}
\end{eqnarray}
Equation \ref{equa:Gamma} is a concise and elegant interim result for the time-varying circulation, where only the coefficient $c_0(t)$ is required. Furthermore, all $c_n(t)$ are functions of $b_n(t)$ which themselves vary only in time. Thus, the spatial variable $x$ or $\Theta$ is eliminated. The formulation in equation \ref{equa:Gamma} leads to an integro-differential equation for $\Gamma(t)$, where according to \citet{isaacs1946airfoil} and \citet{vanderwall1992forschungsbericht}:
\begin{eqnarray}
b_n(t) &=&-\frac{2}{\pi c}\int^\infty_0 \dot{Q}(W(t)-\Lambda)\frac{\left[1+2\Lambda/c-\sqrt{(1+2\Lambda/c)^2-1}\right]^n}{\sqrt{(1+2\Lambda/c)^2-1}}d\Lambda\label{equa:bn}
\end{eqnarray}
\noindent with $\Lambda = W(t)-W(\tau^*)$ and $\dot{Q}(W(t)-\Lambda) = \dot{\Gamma}(t-T) = \dot{\Gamma}(\tau^*)$.
Under the assumption that all explicit variables, for example $u_s(t)$ and $\alpha(t)$, are periodic in time with the angular frequency $\omega=2\pi f$ and the transient starting process is ignored, the resulting circulation $\Gamma(t)$ is periodic in time as well. Thus, the time-derivation of the circulation, which is equivalent to the shed wake vorticity strength, is expressed as:
\begin{eqnarray}
\dot{Q}(W(t)-\Lambda)&=& \sum_{m=-\infty}^{\infty}a_m im\frac{\omega}{{u}_s}e^{im \omega(W(t)-\Lambda)/{u}_s}\label{equa:Q}
\end{eqnarray}
Since $W(t)$ describes the distance travelled by the airfoil through the unsteady inflow, it is the time integral of the free-steam velocity $u_s(t)$. The most general formulation of the coefficients $a_m$ is given by \cite{vanderwall1992forschungsbericht} [the coefficient $imkc/2$ in equation B.41 in \cite{vanderwall1992forschungsbericht} contains a typographical error, the coefficient in front of the integral must be $imk2/c$. This typographical error has no consequence for the rest of \cite{vanderwall1992forschungsbericht}], where: 
\begin{eqnarray}
a_m &=&\frac{A_m}{R_m}\label{equa:am}\\
R_0 &=& 1\\
R_m &=&1+imk\frac{2}{c}\int^\infty_0e^{-im\frac{\omega\Lambda}{{u}_s}}\left(\sqrt{\frac{c}{\Lambda}+1}-1\right)d\Lambda\label{equa:Rm}\\
A_0 &=&\pi c \alpha_s {u}_s \left[\left(1+\frac{\sigma^2}{2}\right)\bar{\alpha}_0+\sigma\left(\bar{\alpha}_{1s}-\frac{k}{4}(1-2a)\bar{\alpha}_{1c}\right)\right]\\
A_m &=& \frac{i^m}{m}\pi c\alpha_s {u}_s (H_{m}(m\sigma)+iH^\prime_{m}(m\sigma))
\end{eqnarray}
An oscillating free-stream and unsteady pitching motion are included in the above coefficients, which represent a closed-form solution of the unsteady wake. By means of the known shed vorticity, the circulation of the wing section is now determined in equation \ref{equa:Gamma}. For this purpose, only the two coefficients $c_0(t)$ and $c_1(t)$ need to be evaluated for the desired unsteady motion as described in \citet{isaacs1946airfoil} and \citet{vanderwall1992forschungsbericht}. 

\subsection{Calculation of the Unsteady Bound Vortex Sheet} \label{sec:vortex-sheet}
The starting point for the new derivation of the unsteady bound vortex sheet is 
\begin{eqnarray}
\gamma_b(\Theta,t) &=&\frac{c_0(t)+\sum^\infty_{n=1}c_n(t)\cos(n\Theta)}{\sin(\Theta)}  \label{equa:gammab}
\end{eqnarray}
\noindent where all of the coefficients $c_n$ are required (\cite{strangfeld2015active}). Although the limits of the unsteady bound vortex sheet are consistent with the steady case \citep{strangfeld2014airfoil}, the time-varying coefficients $c_n$ are still unknown in equation \ref{equa:gammab}.  Equation \ref{equa:bn}, combined with the periodic formulation of the unsteady shed vorticity in equation \ref{equa:Q}, leads to: 
\begin{eqnarray}
b_n(t) &=& -\frac{2}{\pi c}\int^{\infty}_0 \sum_{m=-\infty}^\infty a_m i m \frac{\omega}{{u}_s}e^{im \omega (W(t)-\Lambda)/{{u}_s}} \nonumber\\
&&~~~~~~\cdot \frac{\left[1+2\Lambda/c-\sqrt{(1+2\Lambda/c)^2-1}\right]^n}{\sqrt{(1+2\Lambda/c)^2-1}}d\Lambda \label{equa:b_n1}
\end{eqnarray}
\noindent where equation \ref{equa:surging}, with  $\phi = \omega t$, is time-integrated to obtain $W(t) = \int u_s(t)dt = {u}_s(t-\frac{\sigma}{\omega}\cos(\omega t))$. Furthermore, equation \ref{equa:b_n1} is rearranged and the definition of the reduced frequency $k$ is used to obtain

\begin{eqnarray}
b_n(t) &=& -\frac{2}{\pi c} \sum_{m=-\infty}^\infty a_m i m \frac{2k}{c}e^{im(\omega t-\sigma \cos(\omega t))} \nonumber\\
&&~~~~~~\cdot\int^{\infty}_0 e^{-im\frac{2k}{c}\Lambda} \frac{\left[1+2\Lambda/c-\sqrt{(1+2\Lambda/c)^2-1}\right]^n}{\sqrt{(1+2\Lambda/c)^2-1}}d\Lambda\label{equa:b_n3}
\end{eqnarray}
The coefficient $a_m$ is already determined in general in the equations \ref{equa:am} and $S_m = im (2k/c) \exp [{im(\phi-\sigma\cos(\phi))]}$ is introduced. The substitution $\Lambda = c\tilde{\Lambda}$ with $d\Lambda = c d\tilde{\Lambda}$ simplifies the final equation and the phase angle $\phi = \omega t$ is introduced for simplification, without loss of generality, to give 
\begin{eqnarray}
b_n(\phi) &=& -\frac{2}{\pi} \sum_{m=-\infty}^\infty A_m S_m \frac{\int^{\infty}_0 e^{-imk2\tilde{\Lambda}} \frac{\left[1+2\tilde{\Lambda}-2\sqrt{\tilde{\Lambda}^2+\tilde{\Lambda}}\right]^n}{2\sqrt{\tilde{\Lambda}^2+\tilde{\Lambda}}}d\tilde{\Lambda}}{1+2imk\int^\infty_0e^{-imk2\tilde{\Lambda}}\left(\sqrt{\frac{1}{\tilde{\Lambda}}+1}-1\right)d\tilde{\Lambda}}\label{equa:final}
\end{eqnarray}
Now, the problem of the unsteady bound vortex sheet $\gamma_b(\Theta,t)$ is completely solved using the  integrals given in Appendix of \cite{strangfeld2016airfoil}. By means of the known coefficients $b_n$, $\gamma_b(\Theta,t)$ is determined for all arbitrary amplitudes $\sigma$ and reduced frequencies $k$. Although $A_m$, $S_m$, and the denominator in equation \ref{equa:final} are independent of $n$, the integral in the numerator possesses $n$ as an exponent. Thus, for all desired wave numbers $m$ and coefficients $b_n$, this equation has to be solved separately. This results in excessive processing time because several hyperbolic Bessel functions $K$ and confluent hypergeometric Kummer functions $M$ are part of the solution. Exact solutions of the integrals are given in \cite{strangfeld2016airfoil} for arbitrary $\sigma$ and $k$. Finally, this explicit formulation of the bound vortex sheet enables determination of the two contributions to lift, namely the Joukowsky and impulsive-pressure contributions, expressed below as coefficients, namely (\cite{vanderwall1992forschungsbericht}):

\begin{eqnarray}
C_{L,j}(t) &=& \frac{2}{u(t)} \int^{0.5}_{-0.5}\gamma_b(\bar{x},t)d\bar{x}
\label{equa:Joukowsky}
\end{eqnarray}
\noindent and

\begin{eqnarray}
C_{L,i}(t) &=& \frac{2c}{u^2(t)} \frac{d}{dt}\int^{0.5}_{-0.5}\gamma_b(\bar{x},t)(0.5-\bar{x})d\bar{x} \label{equa:impulsive}
\end{eqnarray}
Equations \ref{equa:Joukowsky} and \ref{equa:impulsive} are particularly useful for evaluating experimental data where the bound vortex sheet strength can be measured directly. They are also useful for identifying the source of deviations of experimental data from theory.  

\section{Experimental Setup}\label{sec:experimentalsetup}
\begin{figure}
\centerline{\includegraphics[width=380px]{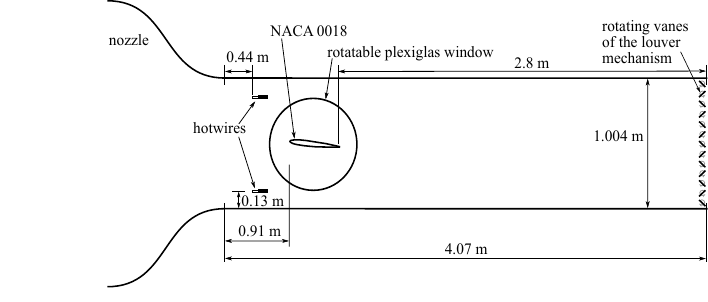}}
\caption{Schematic of the the wind tunnel setup. The airfoil is rotated about its quarter-chord location and surging is achieved by the periodically opening and partially closing the louver vanes at the downstream end of the test section. The unsteady free-stream velocity is recorded upstream of the airfoil by means of two hot-wire probes mounted near the floor and ceiling of the tunnel.}
\label{fig:windtunnelsetup}
\end{figure}

Experiments were performed employing a NACA 0018 airfoil model in the Technion's unsteady low-speed wind tunnel \citep{greenblatt2016unsteady}. It is driven by a speed-controlled 75~kW centrifugal blower, with an 8:1 nozzle contraction ratio and a 0.61~m $\times$ 1.004~m test section area. The maximum free-stream velocity $u_s$ is 55~m/s with a turbulence level of less than 0.1~\%. The ceiling, floor, and side walls of the test section are constructed from Plexiglas to enable maximum optical access for PIV measurements (see Figure \ref{fig:windtunnelsetup}). The tunnel exit is equipped with 13 fully rotatable  louver vanes,  driven by a 0.75~kW servo motor, that control the tunnel surging flow with a maximum area blockage of 95~\%. The tunnel frequency bandwidth, or cutoff, was determined theoretically and experimentally to be:
\begin{eqnarray}
{{f}_{c}}=2\pi {{({{A}}/{{\bar{A}}_{e}})}^{2}}{{{u}}_{s}}/{{L}_{ts}}
\label{equa:freq-cutoff}
\end{eqnarray}
where ${{A}}/{{\bar{A}}_{e}}$ is the ratio of the test-section area to the mean open exit area and ${{L}_{ts}} = 4.07$~m is the test section length. A major advantage of this tunnel is that the cutoff frequency scales linearly with the mean tunnel speed. This means that changes to the test frequency have no effect on the relative tunnel surging amplitude ($\Delta u/{u}_{s}$).


The airfoil ($c=0.348$~m) was mounted rigidly at the vertical center of the test section, between two rotatable Plexiglas windows, each with a diameter of 0.93~m. The pitch axis was located at the quarter-chord point and the leading-edge was positioned $0.825$~m downstream of the nozzle. Synchronous dynamic pitching was achieved using a 1.5kW servo-motor, located above the test section, via a 1:150 belt system attached to the windows. The instantaneous angle-of-attack was measured independently using an optical encoder, where the difference between the commanded and measured values never exceeded $\pm 0.2^{\circ}$. The unsteady free-stream velocity in the test section was measured by averaging the signals of two hot-wire probes, mounted above and below the airfoil, attached to an A.A. Lab Systems Anemometer (AN-1003 Test Module). Calibrations were performed with a Pitot-static (Prandtl) tube and Dwyer Manganese pressure transducer. The airfoil was equipped with 40 symmetrically disposed pressure ports (0.8mm diameter), designated $p_l$ ($l=1,...,40$), and coupled to two piezoresistive pressure scanners (SP-32HD, TE Connectivity) via 44cm long tubes. The pressure lag and amplitude attenuation was found to be negligible for the maximum oscillation frequencies of 1.2Hz considered here \citep[cf.][]{greenblatt2001unsteady,nagib2001effective}. The data acquisition of the surface pressures and the wind tunnel speed were synchronised, acquired at a frequency of 497~Hz, and phase-averaged to obtain $p_l(x,\phi)$ and $u(\phi)$. Wind tunnel blockage corrections were not implemented, due to the low angles-of-attack $(\alpha_{\textrm{max}}=4^{\circ})$ considered and resulting low maximum blockage ratio of 8.7~\%. Furthermore, lift coefficient ratios (see equation \ref{equa:clvanderwall}) vortex sheet differences and form-drag coefficient ratios were considered here, which eliminates any potential wind tunnel bias.

The measured static pressure, which acts normal to the surface, was weighted by the half distance to the neighbouring pressure taps and transformed in the coordinate system of the wing chord. Summations were then employed to obtain the lift force and the form-drag force. The cross product of the static pressure at each pressure port and the distance to the quarter chord was used to obtain the pitching moment. It was recently recognised by \cite{greenblatt2023laminar} that, because surging flows produce streamwise temporal pressure gradients, the local measured static pressure in the wind tunnel  $p_{st} = p_{st}(x, \phi)$ must be used to calculate the surface pressure coefficients $C_{p}(x, \phi)$. Assuming incompressible flow and harmonic surging, \cite{greenblatt2023laminar} showed that the pressure coefficients must be calculated according to 
\begin{eqnarray}
C_{p,l}(x', \phi) = \frac{p_l(x', \phi) - {p_{st}}(0,\phi)} {q (\phi)} + 4 \sigma k \hat{x} \frac{\cos \phi}{(1+\sigma \sin \phi )^2}
\label{correct-cp}
\end{eqnarray}
where $x'$ is measured from the airfoil leading-edge,  ${q}(\phi) = {\raise0.5ex\hbox{$\scriptstyle 1$} \kern-0.1em/\kern-0.15em \lower0.25ex\hbox{$\scriptstyle 2$}}\rho u^{2} (\phi)$ is the tunnel dynamic pressure and $\hat{x}=x'/c$. The correction term in equation \ref{correct-cp} (second term on the right) brings about either no difference $(\alpha = 0^{\circ})$ or near negligible differences $(\alpha \neq 0^{\circ})$ to $C_l$ and $C_m$ due to the correction being applied on both surfaces. However, the correction term brings about enormous changes to the form drag coefficient $C_{dp}$, which otherwise exhibits large non-physical positive and negative oscillations.


\section{Results \& Discussion}\label{sec:results}
\subsection{Surging \& Pitching Wind Tunnel Conditions} \label{sec:WTSurgePitch}
\begin{figure}
\centerline{\includegraphics[width=400px]{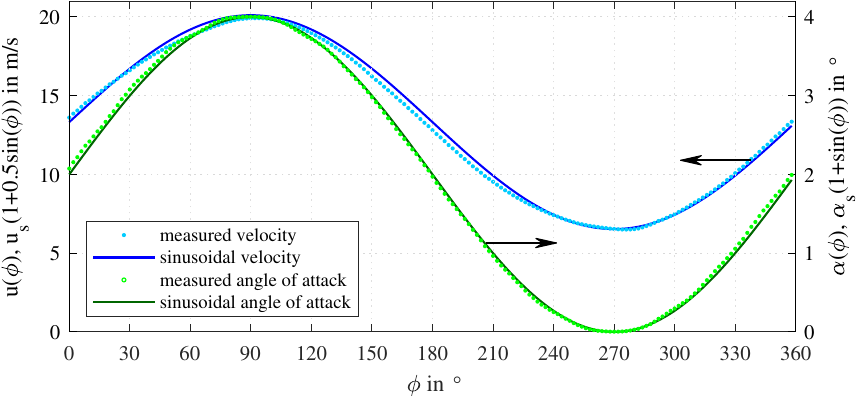}}
\caption{Comparison of the measured free-stream velocity (blue dots) and measured angle-of-attack (green dots) with the sinusoidal functions $u(\phi)=(1+0.51\sin(\phi))13.32$m/s and $\alpha(\phi)=2^\circ+2^\circ\sin(\phi)$ depicted by solid lines.}
\label{velocity_AOA_profiles6790}
\end{figure}

An example of synchronous in-phase ($\tau = 0^{\circ}$) free-stream velocity and angle-of-attack measurements is shown in Figure \ref{velocity_AOA_profiles6790} for ${u}_s = 13.3$ m/s, $\sigma = 0.51$,  $\alpha(\phi)=2^\circ+2^\circ\sin(\phi)$ and $k=0.097$. Free-stream velocities are based on the phase-averaged hot-wire measurements near the wall and ceiling and angles-of-attack are phase-averaged shaft encoder data. Each phase-averaged data set consists of at least 150 periods of the unsteady cycle, where averaging is performed at $\phi$ steps of $2^\circ$ with a window size of $\pm1^\circ$. In addition, each of the data sets is represented by an ideal sine wave determined from a least squares approximation, which serves two purposes. First, it provides representative approximations for the theoretical calculations and second, it allows us to quantify experimental deviations from idealized conditions. The largest detectable deviations are around phase angles $\phi=60^\circ$ and $\phi=180^\circ$, with a 2.5~\% maximum relative error between the measured free-stream and the idealized sine function \citep[also see][]{strangfeld2014airfoil}. The maximum and the minimum exhibit small phase-lags of approximately 4$^\circ$ while the crossing of the steady free-stream velocity shows a phase lead of approximately $-4^\circ$.  The measured angle-of-attack data produced $\alpha_s=2.00^\circ$ and $\alpha_{a}=2.01^\circ$ and the measured angle-of-attack corresponded well with a least-squares sine-wave. Furthermore, a computed cross-correlation between the measured angle-of-attack and the measured velocity profile exhibited a phase lag of 0.0$^\circ$. Finally, hot-wire probes mounted at $1.1c$ upstream and downstream of the leading- and trailing-edges, respectively, indicated that, at the test frequency $f = 1.18$ Hz, phase-lag due to compressibility effects was small, i.e., $\Delta \phi \leq 1^{\circ}$. Thus, the incompressibility condition was not violated.

\subsection{Pure Surging \& Pure Pitching}
\begin{figure}
\centerline{\includegraphics[width=400px]{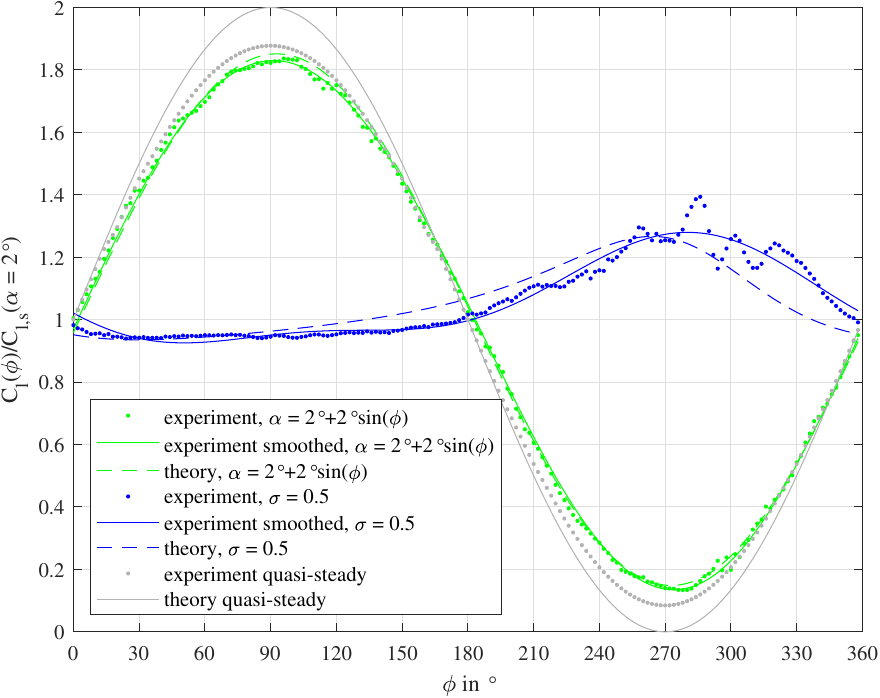}}
\caption{Comparison of experiment and theory for the ratio of unsteady to quasi-steady lift coefficients under pure surging \cite[theory of][]{isaacs1945airfoil} with $\sigma=0.51$ and under pure pitching \cite[theory of][]{theodorsen1935general} with $\alpha(\phi)=2^\circ+2^\circ\sin(\phi)$ at ${Re_s}=3.0 \times 10^{5}$ and $k=0.097$.}
\label{fig:vanderWall_theodorsen_isaacs6790}
\end{figure} 

Quasi-steady and unsteady lift coefficients, relative to their average namely $C_l(\phi)/C_{l,s}(\alpha=2^{\circ})$ are shown for both pure pitching and pure surging experiments (dots), together with corresponding theory (dashed lines), in figure \ref{fig:vanderWall_theodorsen_isaacs6790}. Quasi-steady pitching data at ${Re_s}=3.0 \times 10^{5}$ were generated by pitching the airfoil at $k=0.001$, i.e., two orders of magnitude slower than the unsteady case.  Surprisingly, the quasi-steady pitching data does not correspond to elementary steady theory, and there are two reasons for this. First, the geometric zero angle-of-attack $\alpha=0^{\circ}$ produces a small offset, namely $C_l(\alpha=0^{\circ})= 0.019$ due to a slight asymmetry caused by the slots. Second, the lift-slope is less than the theoretical value, namely $dC_l/d\alpha \approx (0.91)2 \pi$, due to a combination of the relatively thick airfoil and relatively low Reynolds number. Hence, the quasi-steady $C_l(\phi)/C_{l,s}(\alpha=2^{\circ})$ data presented in figure \ref{fig:vanderWall_theodorsen_isaacs6790} does not go to zero when $\alpha=0^{\circ}$ and does not equal 2 when $\alpha=4^{\circ}$.

For the pure unsteady pitching case, the correspondence between the data and theory of \cite{theodorsen1935general}, indicated by the green dashed line, is excellent. Additionally, the data were filtered by summing the first two terms of their Fourier series (solid green line). The excellent correspondence is somewhat misleading because the \emph{changes} between the quasi-steady and unsteady theories are larger than those observed in the experiments. In both theory and experiments, the respectively low and high lift coefficients at $\phi \approx 90^\circ$ and $270^\circ$, relative to quasi-steady theory, are principally due to wake circulation effects \citep{motta2015influence}.  It is likely that the effect observed in the experiments is smaller due to the finite boundary layer thicknesses at the trailing-edge. The data shown in the figure is sometimes presented as counter-clockwise (phase-lag) $C_l$ versus $\alpha$,  hysteresis loops, but the present phase-dependent representation is clearer. Non-circulatory apparent mass effects are not a significant factor, because they only become important at much higher reduced frequencies. Theoretically, at $k = 0.144$ a phase inversion occurs, meaning that the hysteresis loops switch from counter-clockwise to clockwise. \cite{motta2015influence} also showed theoretically, that the phase inversion increases with the airfoil thickness because the potential difference across the airfoil is affected by airfoil finite thickness. In particular, at reduced frequencies less than the phase-inversion point, $d \alpha /dt > 0$ brings about an an increment of the kinematic angle-of-attack, which increases the phase-lag to produce wider counter-clockwise hysteresis loops. Nevertheless, theoretical differences between a flat plate airfoil and a NACA 0018 airfoil at $k=0.097$ are relatively small, namely $\Delta C_l \approx 0.01$.

Quasi-steady surging data were generated by performing quasi-steady pitching experiments, described above, at 11 free-stream velocities between $(1-\sigma)u_s$ and $(1+\sigma)u_s$ and then extracting the data points at $\alpha=2^{\circ}$. Then data corresponding to unsteady free-stream velocities were interpolated from the constant velocity data. Thus, Reynolds number effects, which were relatively minor, were implicitly accounted for in the $C_l(\phi)/C_{l,s}(\alpha=2^{\circ})$ unsteady surging results presented. These data were also filtered by summing the first two terms of their Fourier series (solid line) and the pure unsteady surging theoretical result of  \cite{isaacs1945airfoil} is shown as a dashed line. Overall, that data captures the phase-lag associated with a delay in the development of lift as well as the slightly lower lift coefficient at maximum velocity predicted by \citet{isaacs1945airfoil}. However, the data lags the theoretical result during the deceleration phase ($du/dt<0$), where there is a lift overshoot. This may be due to airfoil thickness effects analogously to the analysis of \cite{motta2015influence}, described above, where increasing airfoil thickness increases phase-lag when compared to flat plate theory. More likely, it is an effect of the finite boundary layer thicknesses at the trailing-edge. Unlike the pure pitching case, surging produces relatively high-frequency oscillations at the beginning of the accelerations phase, i.e., following $\phi > 270^{\circ}$. \cite{greenblatt2023laminar} showed that this was due to bubble-bursting, surprisingly occurring during early stages of the acceleration $(\partial p /\partial x <0)$, because the favorable pressure gradient rapidly drives upper and lower surface separation bubbles aft, rendering them unable to reattach to the airfoil surfaces. 

\subsection{In-Phase Surging \& Pitching}
\begin{figure}
\centerline{\includegraphics[width=400px]{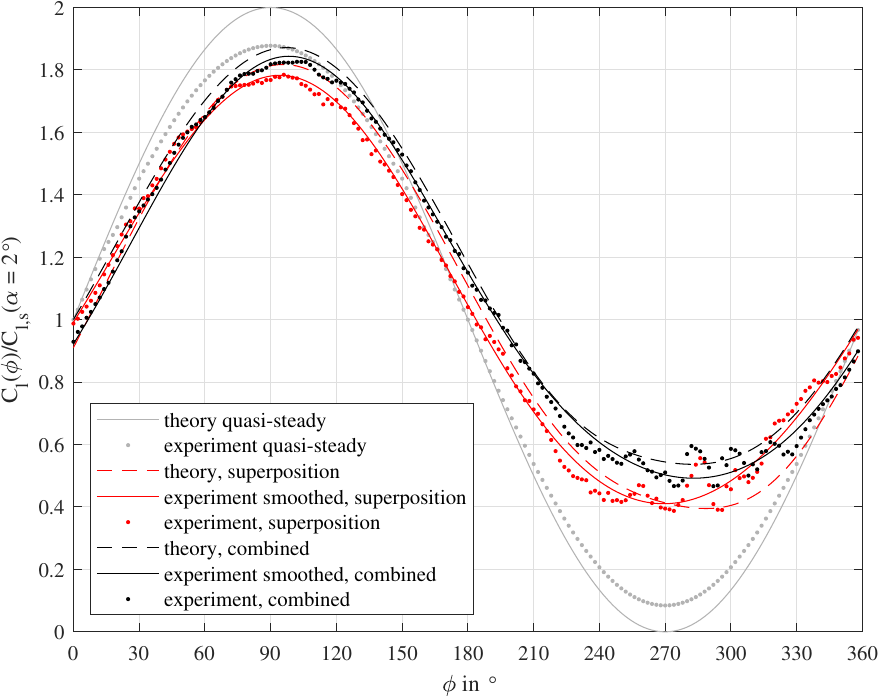}}
\caption{Comparison of experiment and theory for the ratio of unsteady to quasi-steady lift coefficients under synchronous surging-and-pitching with $\sigma=0.51$, $\alpha(\phi)=2^\circ+2^\circ\sin(\phi)$ and $\tau = 0^\circ$ at ${Re_s}=3.0 \times 10^{5}$ and $k=0.097$. Unsteady theory due to \cite{vanderwall1992forschungsbericht} and superposition of the theories of \cite{theodorsen1935general} and \cite{isaacs1945airfoil}.}

\label{fig:vanderWall_theodorsen_isaacs6790_superposition}
\end{figure}

The lift force $L(\phi)$ acting on the airfoil is proportional to both dynamic pressure $q(\phi)$ and angle-of-attack  $\alpha(\phi)$. Therefore we can expect to observe the greatest non-linear effects when pitching and surging are in phase, i.e., at $\tau = 0$. Synchronous in-phase surging-and-pitching results for the conditions described in the previous section, i.e., ${Re_s}=3.0 \times 10^5$ and $k=0.097$, are shown in figure \ref{fig:vanderWall_theodorsen_isaacs6790}. The close correspondence between the data and van der Wall's theory, suggests that the theory adequately captures the dominant unsteady effects of synchronous surging-and-pitching. Note, furthermore, that the effect on the lift coefficient due to synchronous surging-and-pitching is not merely the superposition of their individual contributions. This is evident by superimposing the results of \cite{isaacs1945airfoil} and \cite{theodorsen1935general} and shown as the red line in figure \ref{fig:vanderWall_theodorsen_isaacs6790}, which falls below the theoretical prediction of \cite{vanderwall1992forschungsbericht}. An identical superposition exercise was performed using the individual experimental data data sets, shown as the red dots, and a similar result was obtained. The difference between synchronous surging-and-pitching and the superimposed result, is due to variation in the phase dependant ``effective reduced frequency'' experienced by the airfoil during pitching. For superposition, the reduced frequency is simply $k=\omega c/ 2 u_s$, but with synchronisation it is $k_e(\phi)=\omega c/ 2 u(\phi)$, e.g., 1.5 times lower and higher at $\phi = 90^{\circ}$ and $270^{\circ}$, respectively. Note, furthermore, that the high-frequency lift oscillation, caused by bubble-bursting, is also present during combined surging-and-pitching. This suggests that the bubble-bursting mechanism is not materially affected when the the free-stream velocity and angle-of-attack are in phase. 

\begin{figure}
\centerline{\includegraphics[width=400px]{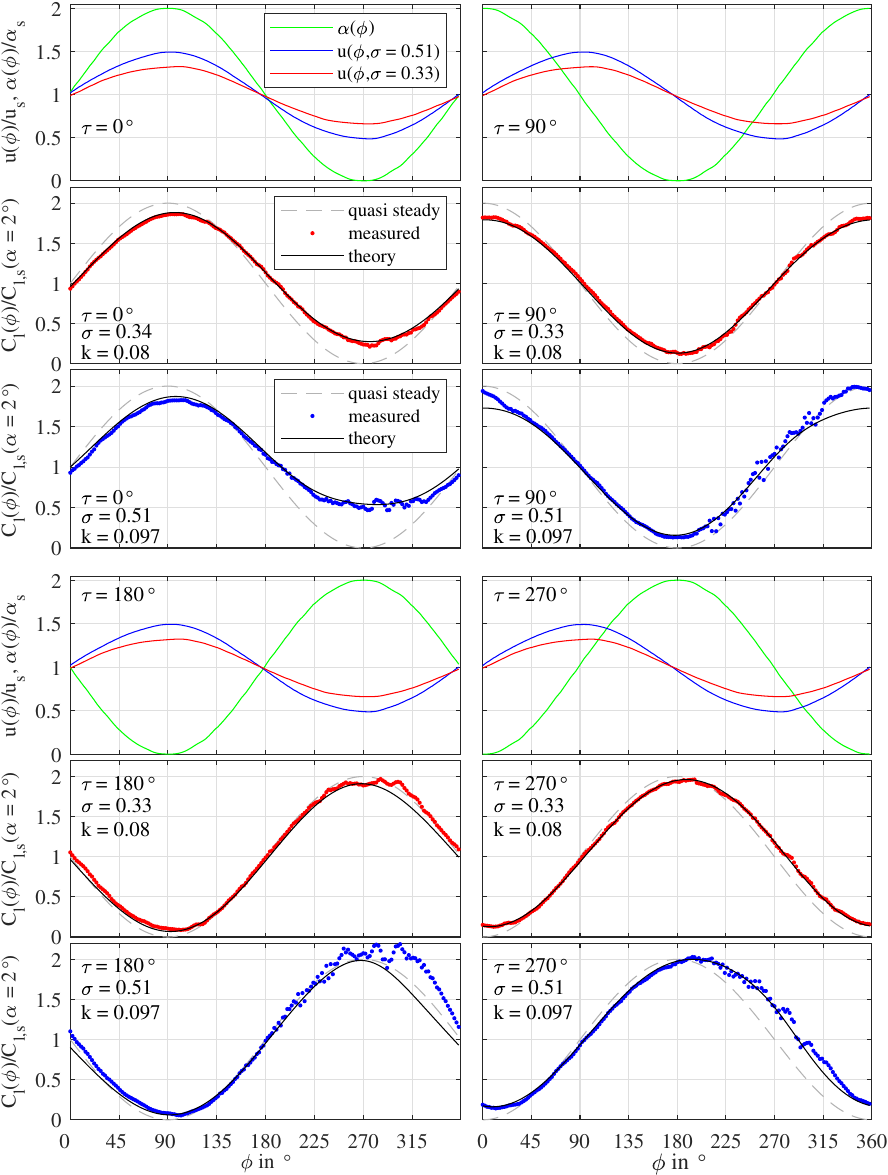}}
\caption{Measured and theoretical unsteady lift coefficient ratio under synchronous surging-and-pitching at $\alpha(\phi)=2^\circ+2^\circ\sin(\phi+\tau)$ at phase-angles $\tau=0^{\circ}$, $90^{\circ}$, $180^{\circ}$ and $270^{\circ}$; red and green symbols correspond to $\{\sigma,k\}=\{0.33,0.08\}$ and $\{0.51,0.097\}$, respectively.}
\label{fig:unsteadylift_multiple_6796}
\end{figure}

\subsection{Effect of Phase-Difference}
\label{sec:phase-difference}
The in-phase pitching-and-surging case with $\sigma =0.51$, described in the previous section, resulted in relatively large effects on the lift coefficient. In this section, we extend the validation to include lower amplitude surging cases $(\sigma=0.33$ at $k=0.08)$ as well as phase-lag angles $\tau = 90^{\circ}$, $180^{\circ}$ and $270^{\circ}$.  Note that the latter two phase differences represent idealized conditions experienced by helicopter rotor-blades in forward flight and vertical axis wind turbine blades, respectively. A summary of the experimental results, together with quasi-steady theory and the theory of \cite{vanderwall1992forschungsbericht} are presented in figure \ref{fig:unsteadylift_multiple_6796}, where four blocks of three vertically-stacked images represent the four phase-differences $\tau$. Measurements of the normalized free-stream velocity $u(\phi)/u_s$ (red and green lines) and angle-of-attack $\alpha(\phi)/\alpha_s$ (blue lines) are shown (see the methods described in section \ref{sec:experimentalsetup}). For the lower amplitude case $(\sigma=0.33$, $k=0.08)$ at $\tau = 0^{\circ}$, the correspondence between data and theory is excellent, much like the high amplitude  $(\sigma=0.51)$ case described in the previous section (lowest image top-left quadrant). The high-frequency oscillations previously observed in the vicinity of $270^{\circ} < \phi <330^{\circ}$ are almost completely eliminated. This indicates a significant change to the bubble-bursting mechanism, which can be directly related to the smaller free-stream oscillation amplitude. First, the temporal pressure gradient is weaker and therefore the separation point does not move as far downstream; and second, the minimum Reynolds number is higher, which also renders the bubbles less likely to burst. 

At $\tau=90^\circ$, the velocity maximum and minimum correspond to peak negative and positive pitch-rates, $(d\alpha / dt)_{\textrm{min}}$ and $(d\alpha / dt)_{\textrm{max}}$, respectively. For the low amplitude case, excellent correspondence is seen between data and theory, with minimal high-frequency oscillations. For the higher-amplitude case, correspondence is excellent for $45^{\circ}<\phi<315^{\circ}$ but a breakdown is observed outside of these bounds. The most likely reason for this is the combination of relatively low Reynolds number combined with the pitch-up motion, that exacerbates bubble-bursting (see section \ref{sec:bubble-bursting}). This hypothesis is reinforced by considering the comparison at $\tau=180^\circ$, where the velocity and angle-of-attack are in anti-phase, i.e., the lowest Reynolds number corresponds with the highest angle-of-attack. In this case, even at low amplitudes, there is a deviation from theory. Another factor that affects the correspondence between experiment and theory is the relative contributions of the Joukowsky and impulsive-pressure components of lift generation, and this is discussed below. The anti-phase $u_s$-$\alpha$ relation is of particular interest, because it is an idealized representation of the blade of a helicopter in forward flight. Therefore the propensity of the bubble to burst under these conditions---which is effectively low angle-of-attack dynamic stall---can have important consequences for rotor performance and noise. For example, the interaction between the turbulent boundary layer with the rotor blade trailing edge discontinuity, typically modelled by the Kutta condition, causes turbulent energy to be scattered as far-field noise in a dipolar pattern \citep{lee2021turbulent}. Therefore, if the Kutta condition is violated by dynamic stall precipitated by bubble-bursting on one or both surfaces, for example during forward flight (cf. $\tau=180^\circ$), this will affect the noise scattering. At $\tau=270^\circ$, the minimum velocity corresponds to the maximum pitch-down rate $(d\alpha / dt)_{\textrm{min}}$ (cf. $(d\alpha / dt)_{\textrm{max}}$ at $\tau=90^\circ$). Here the correspondence improves, most probably because the pitch-down motion is less likely to produce bubble-bursting. 


\begin{figure}
\centerline{\includegraphics[width=400px]{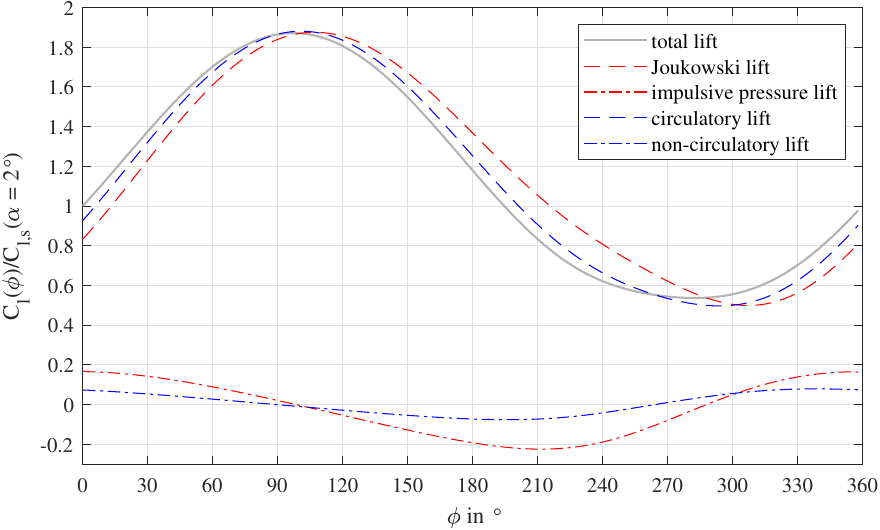}}
\caption{Presentation of the non-dimensionalized individual lift coefficient components as a function of phase-angle for $\tau = 0^\circ$. The sum of the former two and the latter two produce the overall loading developed by Isaacs.}
\label{fig:Joukowsky_circulatory_comparison}
\end{figure}

The theories of \cite{isaacs1945airfoil} and \cite{vanderwall1992forschungsbericht} are limited somewhat in that they provide only circulatory and non-circulatory loads acting on the airfoil. Consequently, they does not provide the explicit vortex sheet distribution $\gamma$; neither do they distinguish between the Joukowsky lift and impulsive-pressure lift contributions. Both of these factors are important for experimental validation of the theory because the vortex sheet strength can be estimated directly from the airfoil pressure measurements and the different contributions to lift generation assist in interpreting the experimental data. Note that overall unsteady loads can be thought of as either the sum of the circulatory and non-circulatory lift components (see eqn. \ref{equa:clvanderwall}) or as the sum of Joukowski lift and the impulsive-pressure lift components (see eqns. \ref{equa:Joukowsky} and \ref{equa:impulsive}). To illustrate this, the bound unsteady vortex sheet was computed for each phase angle under the conditions $\alpha(\phi)=2^\circ+2^\circ\sin(\phi+\tau)$, $\sigma=0.51$ and $k=0.097$, and then integrated numerically to obtain $C_{L,j}$ and $C_{L,i}$. These coefficients were then non-dimensionalized with respect to the steady-state values at $\alpha = 2^\circ$ and presented together with the similarly non-dimensionalized circulatory and non-circulatory lift coefficient components shown in figure \ref{fig:Joukowsky_circulatory_comparison} for  $\tau=0^\circ$ as a function of phase-angle. The difference between circulatory lift and Joukowsky lift is that the latter integration does not explicitly take wake effects into account and is the same integration that would be performed in a steady flow. The difference between the  non-circulatory and impulsive-pressure lift is that the former is proportional to the time-derivative of the free-stream velocity, while the latter is proportional to the time-derivative of the integrated bound vortex sheet weighted by distance from the leading-edge. Note that the sum of both components pairs, i.e., the overall aerodynamic loading predicted by van der Wall are virtually identical for both cases, with maximum differences not exceeding 0.56~\%. These small differences are caused partly by the numerical integration and partly because the Kutta condition cannot be perfectly imposed numerically at the trailing edge (see discussion below).

\begin{figure}
\centerline{\includegraphics[width=400px]{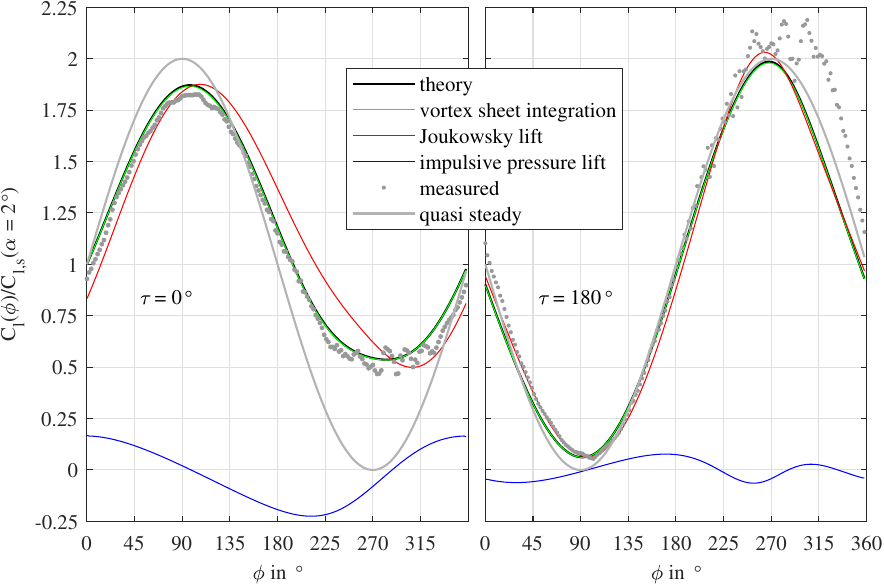}}
\caption{Illustration of the different contributions to unsteady lift, based on the theory of \cite{vanderwall1992forschungsbericht} and integration of the bound unsteady vortex sheet, under synchronous surging $(\sigma = 0.51)$ and pitching ($\alpha(\phi)=2^\circ+2^\circ\sin(\phi+\tau)$) for $\tau=0^{\circ}$ and $180^{\circ}$, at ${Re_s}=3.0 \times 10^{5}$ and $k=0.097$.}
\label{fig:validation_bound_vortex_sheet6794}
\end{figure}

Comparisons between experimental data and theory are shown in figure \ref{fig:validation_bound_vortex_sheet6794} for $\tau=0^\circ$ (on the left) and for  $\tau=180^\circ$ (on the right). The black solid lines are the unsteady lift prediction of van der Wall as described in equation \ref{equa:clvanderwall} and the green solid lines represent the integration of the bound unsteady vortex sheet according to equations \ref{equa:Joukowsky} and \ref{equa:impulsive}. The in-phase and anti-phase results illustrate the relative importance of the Joukowsky and impulsive-pressure lift contributions. In the case of pure surging  described by \cite{strangfeld2016airfoil}, the maximum positive-lift and negative-lift impulsive-pressure component leads and lags the accelerations by approximately $\Delta \phi \approx 30^{\circ}$, respectively. For pure pitching the impulsive-pressure is almost perfectly in phase with the accelerations, due to the relatively small $k\approx 0.1$. Therefore, because the surging-and-pitching impulsive-pressure contributions are almost in phase when  $\tau=0^{\circ}$, their combined contribution to the overall lift is large. In fact, the differences between the quasi-steady and unsteady Joukowsky lift effects and the impulsive-pressure effects are comparable in magnitude (figure \ref{fig:validation_bound_vortex_sheet6794}, left). Now, because bubble-bursting  primarily affects the experimental component of Joukowsky lift, its effect on the overall lift at $\tau=0^{\circ}$ results in relatively small deviations from theory. In contrast, at $\tau=180^{\circ}$, the impulsive-pressure contributions due to pitching-and-surging, whose integrands are weighted by distance from the leading-edge, are close to anti-phase and their net effects are small throughout the cycle. Indeed, (figure \ref{fig:validation_bound_vortex_sheet6794}, right) illustrates that impulsive-pressure lift is negligible, i.e., the Joukowsky lift makes up virtually the entire contribution. By the same reasoning as before, in the corresponding anti-phase experiments, bubble-bursting predominately affects the Joukowsky lift and therefore the integrated lift coefficient data show a relatively large departure from theory.

\subsection{The Bound Unsteady Vortex Sheet}
The close correspondence between the numerically integrated unsteady vortex sheet (equations \ref{equa:Joukowsky} and \ref{equa:impulsive}) and the theoretical unsteady lift prediction (equation \ref{equa:clvanderwall}), means that we can compare the theoretical and experimental bound vortex sheet strengths. Experimentally, the vortex sheet strength is calculated across corresponding (same $x'_l$) high-pressure and low-pressure surface port locations ($l$) according to $\gamma_{b,l} = \Delta p_l / (\rho u(t))$.  For quasi-static representation $\gamma_{b,l,qs}$, the quasi-static pressure differences are used and $u(t)$ is replaced with the corresponding $u_s$. The theoretical unsteady and quasi-steady bound vorticities are calculated according to equations \ref{equa:gammab} and \ref{equa:final} and according to \cite{anderson2011fundamentals}, respectively. Note that the unsteady vortex sheet strength does not tend to zero as $\hat{x} \rightarrow 1$. For both steady and unsteady cases, the  wake vortices induce normal velocities on the vortex sheet and hence the vortex sheet strength must be adjusted such that the normal velocity is negated. In the unsteady case, there are additional normal velocities at the trailing-edge and thus the bound vortex sheet must compensate for this.

\begin{figure}
\centerline{\includegraphics[width=400px]{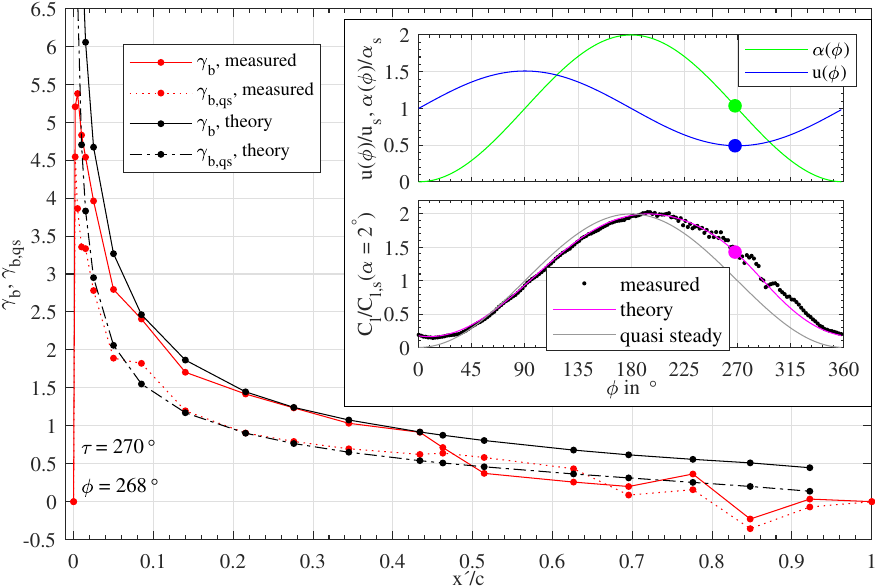}}
\caption{Comparison of measured and theoretical unsteady and quasi-steady bound vortex sheets under synchronous surging $(\sigma = 0.51)$ and pitching at $\alpha(\phi)=2^\circ+2^\circ\sin(\phi+270^\circ)$ for ${Re_s}=3.0 \times 10^{5}$ and $k=0.097$. Inset: free-stream velocity, angle-of-attack and lift coefficient ratios as a function of phase-angle.}
\label{fig:k_mes6796_omega268}
\end{figure}

Vortex sheet comparisons at different phase-lags showed similar results, and hence we only present a representative sample result for the phase angle $\tau=270^\circ$  (see figure \ref{fig:k_mes6796_omega268}) where $\alpha(\phi)$ lags $u(\phi)$ (see upper inset) and unsteady and quasi-steady theoretical results $C_l/C_{l,qs}(\alpha = 2^{\circ})$ are shown (see lower inset). The vortex sheet strengths corresponding to $\phi=268^\circ$ are chosen for comparisons purposes because large differences are observed between unsteady and quasi-steady lift coefficients and the separation bubbles are on the verge of bursting (see indicated points in the insets). Over the airfoil surface defined by $0.05 \leq \hat{x} \leq 0.43$, theoretical unsteady and quasi-steady sheet strengths are very well represented in the experiments. As the leading-edge is approached, from $\hat{x} = 0.05$, significant deviations are observed. This is because the the vortex sheet strengths predicted by a flat plate in potential flow tend to infinity as $\hat{x} \rightarrow 0$  \citep{strangfeld2014airfoil}. Nevertheless, apart from the measurements at $\hat{x}<0.01$, the differences between the experimental and theoretical vortex sheet strengths are comparable. For $\hat{x} > 0.62$ deviations from quasi-steady theory are observed and for and $\hat{x}>0.43$  deviations from unsteady theory are observed. The quasi-steady deviations are due to bubble formation as a result of the  adverse spatial pressure gradients, and is well documented \citep[][and others]{yarusevych2017steady}. The adverse \emph{temporal} pressure gradient in the unsteady case moves the bubble separation point upstream, as observed in pure surging flows by  \cite{greenblatt2023laminar} in the angle-of-attack range $0^{\circ} \leq \alpha \leq 4^{\circ}$. Despite the seeming breakdown in theory due to the presence of separation bubbles,  correspondence between the theoretical and experimental lift coefficient remains strong, and this is true for all phase angles and phase differences, prior to bubble-bursting. Following bubble-bursting at $\phi > 270^\circ$, the correspondence with theory remains strong due to the relative importance of the impulsive-pressure lift component discussed in section  \ref{sec:phase-difference}. 

A limitation of the vortex sheet representation presented here, is that it cannot identify the surface on which the separation bubble has formed, because it is based on the \emph{pressure differences across the airfoil surface}. Consequently, it is also of very little use for examining the mechanisms of bubble-bursting. These factors are addressed in the next section which examines the pressure coefficients on the individual surfaces. 

\subsection{Bubble-bursting \& Form Drag}
\label{sec:bubble-bursting}
\begin{figure}
\centerline{\includegraphics[width=400px]{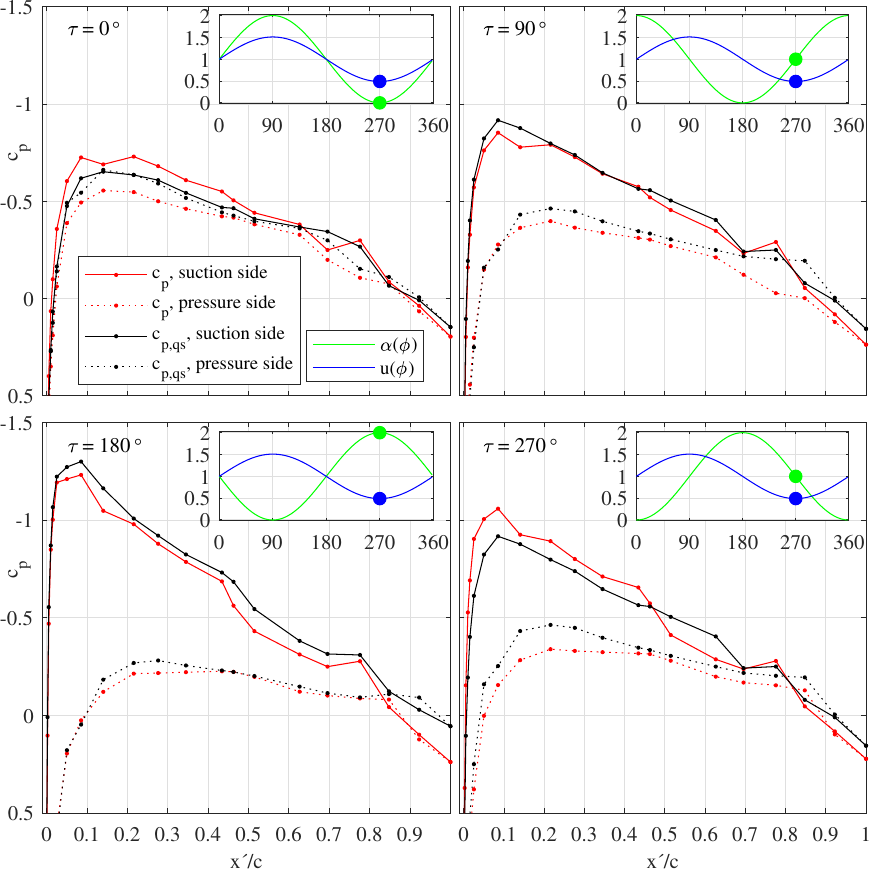}}
\caption{Unsteady and quasi-steady pressure coefficients under synchronous surging-and-pitching at the end of the deceleration phase ($\phi=270^{\circ}$) at four phase-differences.}
\label{fig:Cp_phi=270deg}
\end{figure}

\begin{figure}
\centerline{\includegraphics[width=400px]{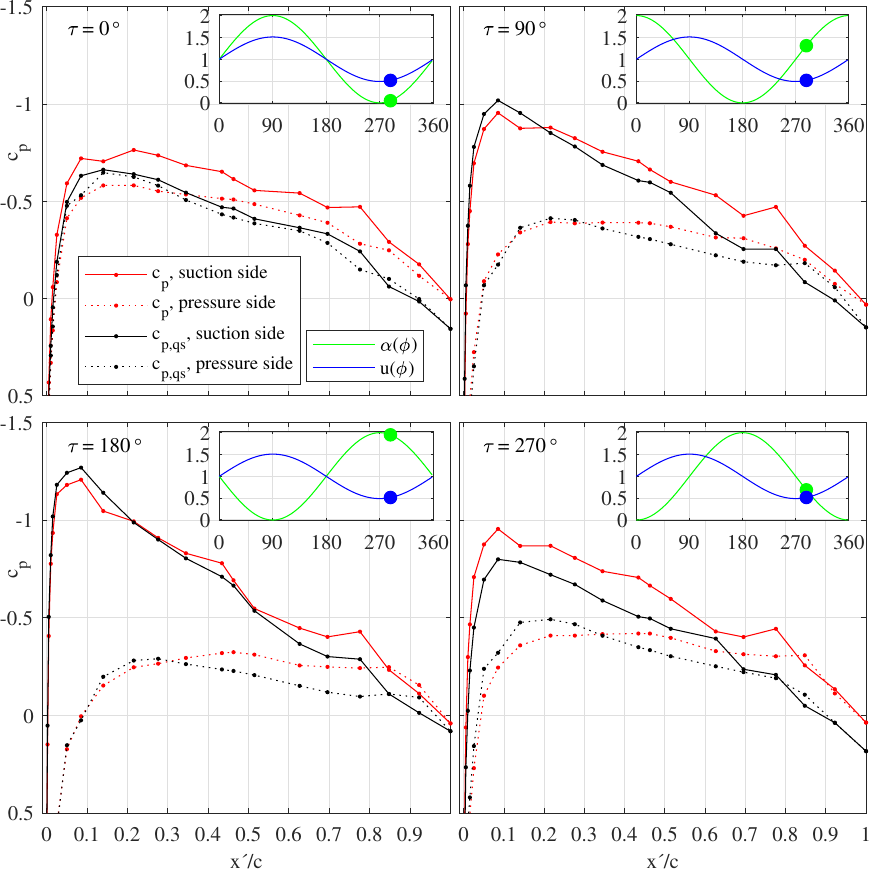}}
\caption{Unsteady and quasi-steady pressure coefficients under synchronous surging-and-pitching during the early stages of the acceleration phase ($\phi=288^{\circ}$) at four phase-differences.}
\label{fig:Cp_phi=288deg}
\end{figure}

Synchronous unsteady and quasi-steady surface pressure coefficient distributions at the end of the deceleration phase ($\phi = 270^\circ$) and the early part of the acceleration phase ($\phi = 288^\circ$) are shown in figures \ref{fig:Cp_phi=270deg} and \ref{fig:Cp_phi=288deg}, respectively, for the four phase-differences $\tau$. Due to a lack of theory for separation bubbles under surging, a comparison between unsteady and quasi-steady data sets is the only viable method presently to isolate unsteady effects by eliminating Reynolds number as a parameter. From figure \ref{fig:Cp_phi=270deg} (end of the acceleration phase), at $\tau=0^{\circ}$, the quasi-steady and unsteady $C_p$ distributions are qualitatively similar, apart from small differences in the region downstream of the leading-edge on both surfaces and a local pressure rise on the upper surface at $\hat{x} \approx 0.7$. The latter indicates a possible unsteady effect on transition within the  separation bubble. At $\tau=90^{\circ}$ and $180^{\circ}$, the main differences are on the \emph{lower surface}, where the initiation of the unsteady local pressure recovery occurs upstream of its quasi-steady counterpart. In contrast, at $\tau=270^{\circ}$, the start of the \emph{upper surface} unsteady pressure recovery is upstream relative to the quasi-steady one.  Therefore, irrespective of the phase-difference, the above observations indicate that the temporal deceleration associated with surging moves transition upstream on either or both of the surfaces, with a overall greater pressure recovery evident at the trailing-edge. This also corresponds to upstream movement of the separation point, discussed below.   

\begin{figure}
\centerline{\includegraphics[width=400px]{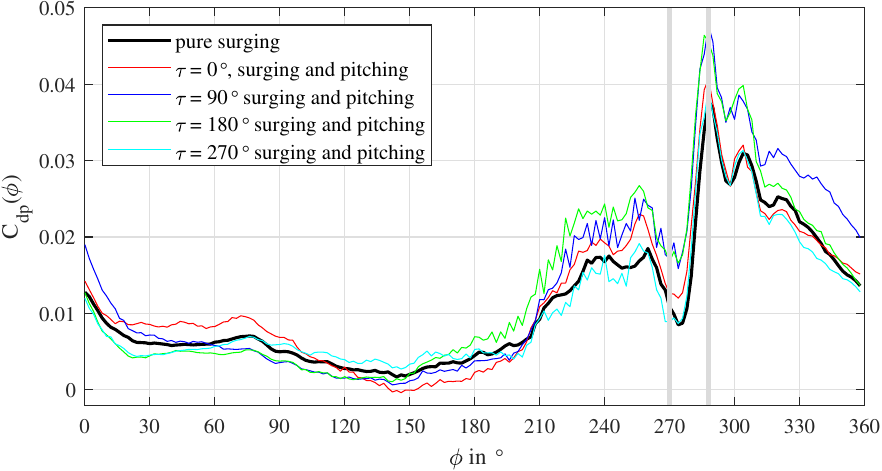}}
\caption{Form-drag as a function of phase angle under pure surging and synchronous surging-and-pitching at four phase differences. Conditions corresponding to $\{\sigma,k\} = \{0.51,0.097\}$ in figure \ref{fig:unsteadylift_multiple_6796}.}
\label{fig:Cdp_summary}
\end{figure}

During the early part of the acceleration phase (see figure \ref{fig:Cp_phi=288deg}), the unsteady pressure distributions at all phase-differences show a significant pressure drop over the aft part of the airfoil, when compared to the steady data. These pressure drops are a clear manifestation of bubble-bursting, irrespective of the corresponding angle-of-attack. It is likely that the lower surface bubble bursts prior to the upper pressure bubble because transition is seen mainly further downstream. Despite the seeming differences in pressure distributions at different phase-differences, the order and mechanism of bubble-bursting is essentially the same and occurs at the same phase angles. This assertion is based on measurements of form-drag variation over the cycle, shown together with the pure surging case at $\alpha = 2^{\circ}$, in  figure \ref{fig:Cdp_summary}, where $\phi = 270^\circ$ and $288^\circ$ are indicated by vertical grey lines. It can clearly be seen that the phases of the local peaks and troughs correspond precisely for all values of $\tau$ and well as for pure surging.

\begin{figure}
\centering
\centerline{\includegraphics[width=6.0in, trim={0 3.5cm 0 2.5cm},clip]{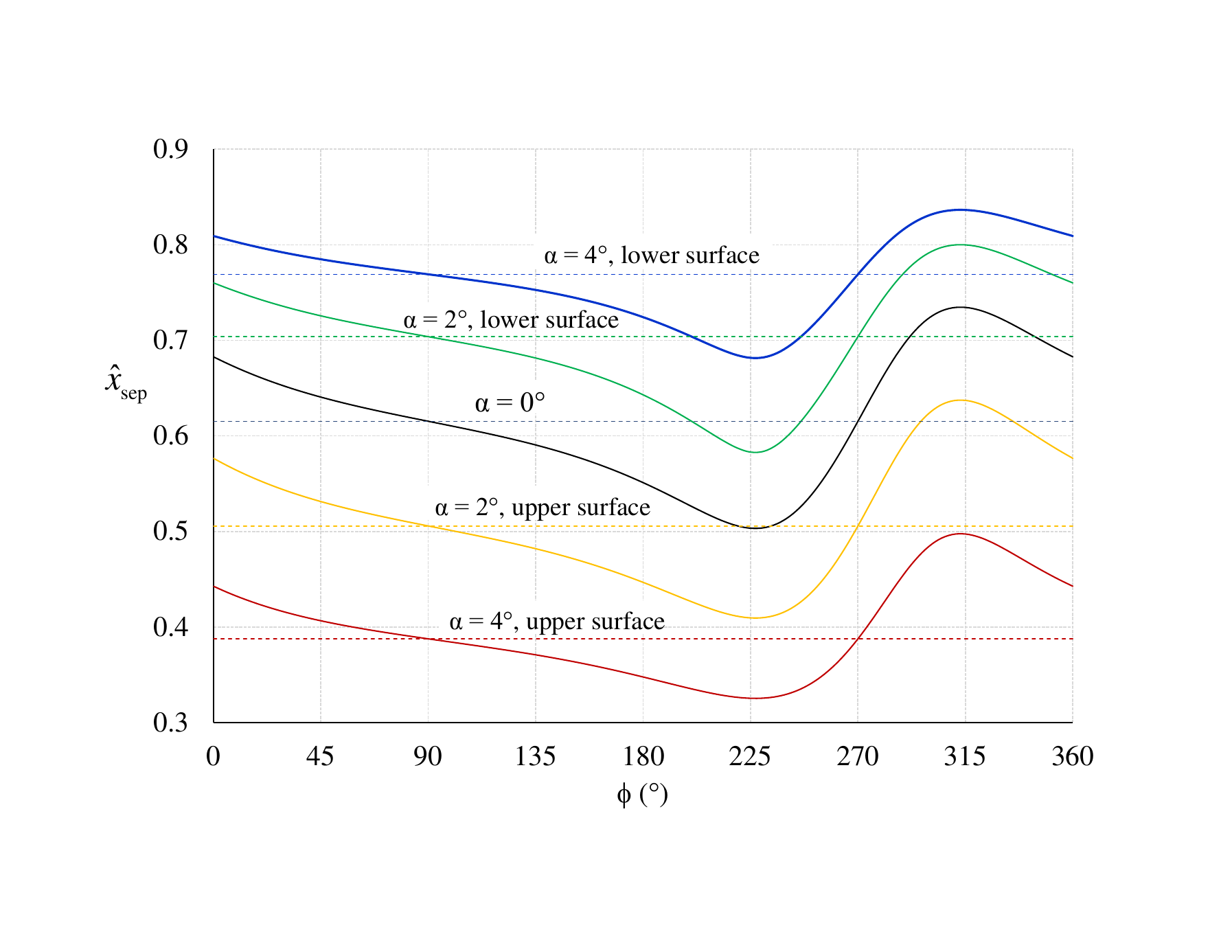}}
\caption{Dimensionless chordwise separation point as a function of phase angle based on quasi-steady boundary layer assumptions; solid lines: quasi-steady momentum integral equation; dashed lines: quasi-steady flow corresponding to $K=0$.}
\label{us-separation}
\end{figure}

The precise phase correspondence between the pure surging flow and synchronous pitching-and-surging at all  phase-differences (figure \ref{fig:Cdp_summary}), suggests that the momentum integral analysis used to determine the movement of the separation point in the former \citep{greenblatt2023laminar} is appropriate here as well.  It is based on the well-known Pohlhausen velocity profile \citep{schlichting2017boundary}, together with the generalized boundary layer parameter \citep{docken1982gust}:
\begin{eqnarray}
K \equiv \frac{\theta^2}{\nu} \left( \frac{{\partial u_e}}{{\partial s}} + \frac{1}{u_e}\frac{{\partial u_e}}{{\partial t}} \right )
\label{Pohl_param}
\end{eqnarray}
\noindent where $\theta$ is the momentum thickness, $u_e$ is the local velocity on the surface $s$ (assumed to be at the edge of the boundary layer) measured from the stagnation point, and $K = - 0.1567$ indicates the separation point. For surging and pitching, the unsteady term on the right hand side of equation \ref{Pohl_param} can be written as:
\begin{eqnarray}
\frac{1}{u_e}\frac{{\partial u_e}}{{\partial t}} = \frac{1}{u}\frac{{\partial u}}{{\partial t}} -\frac{1}{2(1-Cp)} \frac{\partial C_p}{\partial t}
\label{US_Pohl_param}
\end{eqnarray}
\noindent where the first term is due to surging and does not depend on $s$, while the second term is mainly due to pitching and depends on both $s$ and $t$. For the angle-of-attack range considered in this study, the surging term is an order of magnitude greater than the pitching term near the leading-edge, and increases to more than two orders of magnitude further along the chord. We can therefore neglect the pitching term and compute the time-dependent separation point, using a vortex lattice method \cite{drela1989xfoil} to obtain both terms in equation \ref{Pohl_param}. (Strictly speaking, an unsteady vortex lattice method, e.g., \cite{murua2012applications} should be used, but for the $k \sim 0.1$, the differences are negligible.) The separation point, as a function of the phase angle is therefore the same as that obtained at constant angles-of-attack and is shown in figure \ref{us-separation} \citep[see][]{greenblatt2023laminar}. It is seen from the figure that deceleration and acceleration phases produce non-symmetric upstream and downstream movement of the separation separation point, respectively. From  $\phi \approx 230 ^{\circ}$, the predicted separation point begins to move rapidly downstream, crossing its steady value at $\phi = 270 ^{\circ}$, and attaining its furthest downstream  location at $\phi \approx 315 ^{\circ}$. The figure shows furthermore, that the predicted lower surface separation points are further downstream than those on the upper surface, and this is consistent with the  observation that bubble transition occurs further downstream on the lower surface in figure \ref{fig:Cp_phi=288deg}. 

It can be concluded therefore that the unsteady effects on LSB separation and transition are driven predominantly by the surge-induced pressure gradient. It is likely this will also be the case for larger pre-stall angle-of-attack amplitudes. Due to its potential occurrence on the rotating blades of various rotorcraft, this low angle-of-attack dynamic stall mechanism should be studied further both experimentally and computationally. 

\section{Conclusions}
This paper presented a combined theoretical and experimental study of synchronous pitching-and-surging on an airfoil at low pre-stall angles-of-attack $(0{}^\circ \le \alpha \le 4{}^\circ )$.  The theoretical approach was based on the most general formulation of the problem, presented by \cite{vanderwall1994on}, for synchronous pitching-and-surging at four phase-differences, namely $0^{\circ}$, $90^{\circ}$, $180^{\circ}$ and $270^{\circ}$. The theory was then extended to explicitly compute the unsteady bound vortex sheet strength, which facilitated computation of the individual Joukowsky and impulsive-pressure lift components. Experiments were performed by measuring unsteady surface pressures on a NACA 0018 airfoil, in an unsteady wind tunnel at an average Reynolds number of $3.0\times 10^5$. The majority of unsteady pressure data were acquired at free-stream oscillation amplitudes of 51~\%, with an angle-of-attack range of $2^\circ \pm 2^\circ$, and a reduced frequency of 0.097. Quasi-steady surface pressure data were generated by pitching the airfoil at reduced frequencies of 0.001, at eleven nominally constant free-stream velocities encompassing the unsteady range---and then interpolating pressures at corresponding phases within the unsteady cycle. 

Excellent correspondence was obtained between theoretical and experimental lift coefficients for pure pitching throughout the oscillation cycle. Good qualitative correspondence was obtained between pure surging theory and experiments, apart from high-frequency oscillations observed in the experiments during the early stages of the acceleration phase. These oscillations, documented previously by our group, were attributed to bursting of lower and upper surface separation bubbles, which modify the Kutta condition. With synchronous in-phase surging-and-pitching, excellent correspondence was obtained, representing the first direct experimental validation of the general theory of \cite{vanderwall1994on}. It was shown, both theoretically and experimentally, that the lift coefficient cannot be accurately represented by merely superimposition of surging and pitching effects. This is because the effective reduced frequency changes as the free-stream velocity changes, which is not accounted for when pure pitching and pure surging are superimposed. At phase-differences of $90^{\circ}$ and $180^{\circ}$, relatively large deviations from theory were observed during the beginning of the acceleration phase, and these were attributed to the relative impulsive-lift and Joukowsky contributions to the overall lift coefficient. Specifically, at phase differences of $90^{\circ}$ and $180^{\circ}$, the impulsive-lift contributions are relatively small and hence the effect of bubble-bursting, which modifies the Kutta condition, has a dominant effect on the overall lift coefficient. The opposite was true for phase differences of $0^{\circ}$ and $270^{\circ}$

The correspondence between unsteady theoretical and experimental bound vortex sheet strengths was excellent on the upstream half of the airfoil chord. However, the presence of separation bubbles resulted in a breakdown of this correspondence.  Nevertheless, at phase angles where the impulsive-pressure lift component was dominant, i.e., at phase differences of $0^\circ$ and $270^\circ$, the overall lift coefficient correspondence was strong. 

Finally, examination of the airfoil pressure distributions indicated upstream movement of bubble transition on both surfaces, consistent with observations under pure surging. During the early part of the acceleration phase, bubble-bursting was identified irrespective of the phase difference. This was consistent with a boundary layer integral analysis that predicted rapid downstream movement of the lower and upper surface separation points. The bursting produced large form-drag oscillations that occurred at identical phase angles within the oscillation cycle, irrespective of the phase difference between surging and pitching, and fully consistent with observations under pure surging. The bubble-bursting dynamic stall mechanism, observed here at low pre-stall angles of attack, can have  important implications for rotorcraft blade performance and noise emissions. \\

\textbf{Acknowledgements}
The authors are grateful to Professor P. P. Friedmann for bringing their attention to this problem, to Professor B. G. van der Wall for his support and the kind email contacts, and to Dr S. Born for helping to solve the integro–differential equations. This research was supported in part by the Israel Science Foundation (grant no. 840/11) and by a PhD scholarship by ‘Stiftung der deutschen Wirtschaft’.





\bibliographystyle{jfm}
\bibliography{NDT_library}

\end{document}